\documentstyle[prd,aps,psfig,preprint]{revtex}
\tighten
\begin{document}

\title{Accretion of Cold and Hot Dark Matter onto Cosmic String Filaments}

\author{V. Zanchin$^{1, 2}$\footnote[1]{zanchin@het.brown.edu},
J.A.S. Lima$^{1, 3}$\footnote[2]{limajas@het.brown.edu} and
R. Brandenberger$^{1}$\footnote[3]{rhb@het.brown.edu.}}

\smallskip

\address{~\\$^1$Physics Department, Brown University, Providence, RI. 02912,
USA.}

\address{~\\$^2$Departamento de F\'{\i}sica, Universidade Federal de Santa
Maria, \\97119-900, Santa Maria, RS, Brazil.}

\address{~\\$^3$Departamento de F\'{\i}sica Te\'orica e Experimental, \\
     Universidade Federal do Rio Grande do Norte,
     59072 - 970, Natal, RN, Brazil.}

\maketitle

\vskip 1.5cm
\begin{abstract}

\noindent The Zeldovich approximation is applied to study the accretion
of hot and cold dark matter onto moving long strings. It is assumed that such
defects
carry a substantial amount of small-scale
structure, thereby acting gravitationally as a Newtonian line source whose
effects dominate the velocity perturbations.
Analytical expressions for the turn-around surfaces are derived and the mass
inside of these surfaces is calculated. Estimates are given for the redshift
dependence of $\Omega_{nl}$, the fraction of mass in nonlinear objects.
Depending on parameters, it is possible to obtain $\Omega_{nl} = 1$ at the
present time. Even with hot dark matter, the first
nonlinear filamentary structures form at a redshift close to 100, and there is
sufficient nonlinear mass to explain the observed abundance of high redshift
quasars and damped Lyman alpha systems.  These results imply that moving
strings with small-scale structure are the most efficient seeds to produce
massive nonlinear objects in the cosmic string model.

\end{abstract}

\vfill

\setcounter{page}{0}
\thispagestyle{empty}

\vfill

\noindent BROWN-HET-1049 \hfill        June 1996.

\noindent hep-ph/yymmdd \hfill Typeset in REV\TeX

\vfill\eject

\baselineskip 24pt plus 2pt minus 2pt

\section{Introduction}

The cosmic string model may be the most promising alternative to
inflation-inspired theories of structure formation (see Ref. \cite{CSreviews}
for recent reviews). Many predictions of the model, for example a
scale-invariant primordial spectrum of density perturbations\cite{ZV81} and the
amplitude of cosmic microwave background (CMB) anisotropies on large angular
scales\cite{CMBstrings}, are at
least qualitatively in agreement with observations.

The main problem facing the cosmic string model is the computational difficulty
in analyzing its predictions. Cosmic strings will form during a phase
transition in the very early Universe provided that certain well known
topological
criteria\cite{Kibble76} are satisfied by the matter theory. It is also well
established that the network of strings, once formed, will approach a
``scaling" solution\cite{Vil85}, i.e. the distribution of strings will become
independent of time if all lengths are scaled to the Hubble radius.

However, the exact form of the scaling solution is not known. At any time $t$,
there are two components to the string network: loops with radius $R \ll t$,
and ``long" string segments with curvature radius $R_c \sim t$. The best
available numerical simulations
\cite{CSsimuls} indicate that more mass is in the long strings than in loops. A
further uncertainty pertains to the amount of small-scale structure which
builds up on the long string segments as a consequence of the involved dynamics
of the string network. Long straight
strings with no small-scale structure carry no local gravitational potential.
Their only effect is to lead to a conical structure of space perpendicular to
the string\cite{Vil81}. Their transverse velocities will typically be close to
the speed of light since the underlying dynamics is governed by the
relativistic wave equation. Strings carrying small-scale structure, on the
other hand, generate a local Newtonian potential\cite{Carter90} in addition to
the conical distortion of space.

Given the lack of knowledge about the details of the initial string
distribution, it is important to study all mechanisms by which strings can give
rise to structures. Loops will produce roughly spherical objects by
gravitational accretion\cite{ZV81}, long strings without small-scale structure
generate planar overdensities\cite{{SilkVil},{TV86}}, while long strings with
small-scale structure induce - since they act as Newtonian gravitational
line sources - filamentary objects\cite{{VV91},{TV92},{Vollick1}}.

In this paper we will study filament formation, i.e. the accretion of both cold
dark matter (CDM) and hot dark matter (HDM) onto a moving Newtonian line
source,
by means of the Zeldovich approximation\cite{Zel70} and its suitable adaptation
to HDM\cite{PBS90}. The Zeldovich approximation is a Lagrangian perturbation
technique which is an improvement over standard linear perturbation theory
and is widely applied in cosmology.

The accretion of CDM onto static string loops was analyzed a long time
ago\cite{TB86}. Subsequently, it was realized\cite{{BKST87},{VilShafi}} that -
in contrast to theories based on adiabatic perturbations - the cosmic string
model also appears viable if
the dark matter is hot. The effects of string loops on HDM were then analyzed
carefully\cite{{BKST87},{BW88}} by solving the collisionless Boltzmann
equation, i.e. keeping track of the entire phase space distribution of the dark
matter particles. Studies
of nonlinear clustering about static string loops were also performed\cite{SMB}
by means of N-body simulations. At the time when string loops are produced,
they typically have
relativistic velocities. Hence, it is important to study the accretion of dark
matter by moving seeds. Bertschinger\cite{EB87} applied the Zeldovich
approximation to study the case of CDM. He found that the total nonlinear mass
is to a first approximation independent of the velocity $u_i$ of the loop. The
decrease in the width of the nonlinear region is almost exactly compensated by
the
increase in the length of the structure. The study of clustering onto moving
string loops was recently\cite{RM96} generalized to HDM. In this case, the
total nonlinear mass was shown to decrease with increasing $u_i$.
Nevertheless, it was demonstrated that sufficient accretion to form nonlinear
objects by a redshift of $z = 4$ takes place.

Since the string network simulations indicate that long string would be more
important for structure formation than loops, attention turned to the formation
of string wakes. The accretion of CDM was analyzed in detail by means of the
Zeldovich approximation in Ref. \cite{SVBST}. In Ref. \cite{PBS90}, a
modification of the Zeldovich approximation was introduced which made it
possible to study the clustering of HDM in a simple way. This approximation was
tested against the more rigorous approach obtained by solving for the phase
space distribution of dark matter particles by means of the collisionless
Boltzmann equation. The two methods were shown to give very similar results.

As emphasized by Carter\cite{Carter90}, it is likely that long cosmic strings
have a substantial amount of small-scale structure, and that hence the
Newtonian line potential gives rise to more important gravitational effects on
dark matter particles than the velocity perturbation induced by the conical
structure
of space perpendicular to the string. The accretion of cold\cite{Vollick1}
and hot\cite{Vollick2} dark matter onto a moving string with a wiggle has been
analyzed several years ago. However, a better description of the coarse-grained
small-scale structure is by an effective Newtonian gravitational line
source\cite{Carter90,VV91,TV92}. Hence, in this paper we will study the
accretion of CDM and HDM onto a moving Newtonian line source by means of the
Zeldovich and modified Zeldovich approximations, respectively.  We will neglect
the velocity perturbation which is due to the conical structure of space. The
main result is that
even in the case of HDM, nonlinear structures form at a redshift close to 100.
For static line
sources, our results reduce to those of Ref. \cite{Aguirre}. As we shall see,
for reasonable values of the parameters of the cosmic string
model, either with CDM or HDM, the theory is compatible with the present
observational constraints coming from the abundance of high redshift quasars.
Some of our results have been obtained previously in \cite{VV91} and
\cite{TV92}. However, our analysis is more detailed and goes further, and some
of the conclusions concerning the accretion of hot dark matter are different.

The outline of this paper is as follows. In the following section we give a
brief review of the standard and modified Zeldovich approximations. In
Section 3 we study the accretion of cold dark matter onto a moving line source,
and in Section 4 we modify the analysis for hot dark matter. Section 5 concerns
an estimate of the total nonlinear mass as a function of redshift $z$ for the
cosmic string filament model. We conclude
with a discussion of the main results. We work in the context of an expanding
Friedman-Robertson-Walker Universe with scale factor $a(t)$. Newton's
constant is denoted by G, $h$ denotes the Hubble expansion parameter in units
of $100$ km s$^{-1}$ Mpc$^{-1}$, and $\Omega$ is the ratio of energy density to
critical density.

\section{ The Zeldovich Approximation}

The Zeldovich approximation\cite{Zel70} is a first order Lagrangian
perturbation method based on analyzing the trajectory of individual particles
in the presence of an external gravitational source. In an expanding Universe,
the physical position of a dark matter particle is given by

\begin{equation}
{\bf  r}({\bf q},t)  = a(t)\lbrack{{\bf q} - {\mbox {\boldmath $\psi$}}
({\bf q},t)}\rbrack \quad,
\label{eq:01}
\end{equation}
where {\bf q} is the comoving coordinate and {\boldmath $\psi$} is the
comoving displacement due to the gravitational perturbation.

By combining the Newtonian gravitational force equation, the Poisson equation
for the Newtonian gravitational potential, using conservation of matter, and
expanding to first order in {\boldmath $\psi$} one arrives at the following
equation (valid in a matter dominated Universe) for the comoving
displacement of a background particle\cite{Zel70,EB87,RM96}

\begin{equation}
\frac{\partial^{2} {\mbox {\boldmath $\psi$}}}{\partial t^2} +
\frac{2\dot{a}}{a}\frac{\partial{\mbox {\boldmath $\psi$}}}{\partial t}
 +\frac{3\ddot{a}}{a}{\mbox {\boldmath $\psi$}} =
{\bf S}({\bf q},{\bf q}')  \quad,
\label{eq:02}
\end{equation}
where a dot means total time derivative
and {\bf S} stands for the source term.

Up to this point, the analysis is quite general. We now specialize to the case
in which the perturbations are generated by a long straight cosmic string with
small scale structure whose strength is given by the value of $G \lambda$,
where $\lambda = \mu - T$, $\mu$ and $T$ being
the energy per unit length and the tension of the string, respectively. If the
position of the string is denoted by ${\bf  r}'(t)= a(t) {\bf q}'(t)$,
then the source term $S$ is given by:
\begin{equation}
{\bf S}=\frac{2\lambda G}{a^2}\frac{{\bf q}-{\bf q}'}{|{\bf q}-{\bf q}'|^2}
\quad.
\label{eq:03}
\end{equation}

In general, there are two sources of perturbations induced by strings, the
wakes produced by the conical structure of space perpendicular to the string,
and the Newtonian gravitational line source given by (\ref{eq:03}). The effect
of wakes can be modeled by a nonvanishing initial velocity perturbations
towards the plane behind the string. However, since we are interested in
exploring filament formation,
we shall neglect wakes. Hence, the appropriate initial conditions for the
perturbed displacement
are {\boldmath $\psi$}$(t_{s}) = \partial{\mbox {\boldmath $\psi$}}(t_{s})/
\partial t =0$, where $t_s$ is the time when the perturbation is set up
(see below). Using the Green function method,
the  solution of Equation (\ref{eq:02}) may be written in
the following form
\begin{equation}
{\mbox {\boldmath $\psi$}}({\bf q},t)=D_{1}(t){\bf I}({\bf q},t)
- D_{2}(t){\bf J}({\bf q},t) \quad, \label{eq:04}
\end{equation}
with
\begin{eqnarray}
 {\bf I}& = &\int_{t_{s}}^{t}{{\bf S}\frac{D_{2} dt}{W}} \quad,
\label{eq:05}\\
 {\bf J} &=&\int_{t_{s}}^{t}{{\bf S}\frac{D_{1} dt}{W}} \quad.
\label{eq:06}
\end{eqnarray}
Here, $D_{1}$ and $D_{2}$ are the two independent solutions of the homogeneous
equation, and $W$ is the Wronskian. We take $D_{1}$ and $D_{2}$ to be the
growing and decaying modes, respectively. In the case of a spatially flat
(Einstein - de Sitter) Universe (to which we will restrict our attention)
and for $t_i\geq t_{eq}$ we have $a =  (t/t_{0})^{2/3}$, $D_{1}= a$ and
$D_{2} = a^{-3/2}$.

{F}or cold dark matter $t_{s}$ is the true initial time $t_{i}$ (the time when
the string sets up the perturbation),
while for hot dark matter we must take neutrino free streaming
into account. In principle, one should study the evolution of the phase space
density by means of the Boltzmann equation and integrate over momenta to obtain
the density distribution. However, as demonstrated in \cite{PBS90} in the case
of planar accretion, it is a good approximation to use Equation (\ref{eq:04})
but with modified initial conditions. Since the effective perturbations due to
the seed survive only
on scales greater than the free-streaming length \cite{PBS90} $\lambda_{J}
\approx v(t)(z(t) + 1)t$, where $v(t)$ is the r.m.s. velocity of the hot dark
matter particles at time $t$, we can take for HDM a scale-dependent starting
time $t_s$,
namely the time when $|{\bf q} - {\bf q}'| = \lambda_J(t_s(q))$
(where $|{\bf q} - {\bf q}'|$ is the relative
distance between the seed and the dark particle) which yields:
\begin{equation}
t_{s}=t_{s}({\bf q})=\frac{t_0\lambda_{0}^3}{|{\bf q}-{\bf q}'|^{3}} \quad,
\label{eq:07} \end{equation}
where $\lambda_{0}= v_i t_i\sqrt{z_i + 1}$. The implementation of this initial
condition in the basic solutions, in order to account for the effects of free
streaming, is called the {\it modified} Zeldovich approximation\cite{PBS90}.
In the case of static filaments,
the above HDM starting condition is unambiguous. However, for moving filaments
there is an ambiguity. For a fixed ${\bf q}$, the value of $t_s({\bf q})$
changes as the filament moves. In fact, $t_s({\bf q})$ may initially be smaller
than $t$ and later, when the filament is closer to ${\bf q}$, increase to a
value larger than $t$. We will specify how to deal with this ambiguity in
Section 4.

A comment is in order concerning the motion of the line source. If we suppose
the string source is
straight (and infinite), pointing along the $z$ axis, then the acceleration
along the $z$ axis does not affect the motion of the string. That is, we can
interpret ${\bf q}'$ in Equation
(\ref{eq:07}) as a vector position in the perpendicular plane to the string.
The same is also valid for the motion of a fluid dark particle, since its
$z$-motion is not perturbed by the string source. Thus, in the formulae related
to the filament all vectors are to be thought as belonging to the $xy$ plane.
For instance, in Equation (\ref{eq:03}) we have to take ${\bf q} =
(q_{x},q_{y})$ .

The dark matter particles are initially moving away from the string with the
Hubble flow. The gravitational action of the string slows this expansion. At
the time when $d{x}/dt = 0$ the particles decouple from the Hubble flow. This
is called ``turn-around". At any time
$t$, we can compute the value of $q_x$ which is turning around. We denote this
quantity by $q_{nl,x}$. The above turn-around condition, together with Equation
(\ref{eq:01}) yields
\begin{equation}
q_{nl,x}=\frac{a}{\dot a}\dot{\psi_x}({\bf q}_{nl},t)+\psi_x({\bf q}_{nl},t)
\quad, \label{eq:09}
\end{equation}
which determines the comoving turn-around surface.

Using Equations (\ref{eq:04}), (\ref{eq:05}) and (\ref{eq:06}) we get
\begin{eqnarray}
q_{nl,x}&=& 2aI_{x}({\bf q}_{nl},t)+\frac{1}{2}a^{-3/2}J_{x}({\bf q}_{nl},t)\,
, \nonumber\\
    &=&  2\psi_{x}({\bf q}_{nl},t)+\frac{5}{2}a^{-3/2}J_{x}({\bf q}_{nl},t).
\label{eq:10}
\end{eqnarray}

In order to determine the turn-around surface we must integrate Equations
(\ref{eq:05}) and (\ref{eq:06}) for which we have to specify the position of
the string, i.e. ${\bf q}'(t)$.
We choose initial condition for the seed motion in such a way that
${\bf q}'(t_{i})=0$, and $\dot {\bf r}'({t_i})={\bf u}_{i}$.
Since the peculiar velocity of the string decreases as $a(t)^{-1}$, we have
\begin{equation}
{\bf q}'(t) = q_{i}\left[1-\left(\frac{a_i}{a}\right)^{1/2}\right],
\label{eq:11}
\end{equation}
where $q_{i} = 3u_{i}t_{i}/a_{i}$ is the final comoving position of
the string.

\hspace{.3in}

\section{Accretion of Cold Dark Matter}

In this section we study the accretion of CDM onto a moving filament.
By considering the motion of the string along the $y$ direction (then ${\bf q}'
= q'(a) e_{y}$) and using
(\ref{eq:11}), (\ref{eq:03}), (\ref{eq:05}) and  (\ref{eq:06}) we obtain
\begin{eqnarray}
I_{x}&=&\frac{9t_{0}^2}{5}\frac{\lambda G q_{x}}{q_{x}^{2} + (q_{y}-q_{i})^{2}}
\left\lbrace ln\left(\frac{a}{a_s}\right) + ln\left(\frac{q_{x}^{2}+
\left[q_{y}-q'(a)\right]^{2}}{q_{x}^{2}+(q_{y}-q_s)^{2}}\right)+ \right.
                   \nonumber \\
& &\left.
2\frac{q_{y}-q_i}{q_x}\left\lbrack tan^{-1}\left(\frac{q_{y}-q'(a)}{q_x}\right)
 -tan^{-1}\left(\frac{q_y-q_s}{q_x}\right)\right]\right\rbrace \, ,
\label{eq:12}\\
J_{x}&=& \frac{9t_{0}^2}{5}\lambda G q_{x} Y({\bf q},a),
     \label{eq:13}
\end{eqnarray}
where $q_s=q_i(1-\sqrt{a_i/a_s})$ denotes the position of the string at the
time when clustering for HDM begins (for CDM $q_s = 0$), and $Y$
stands for the integral
\begin{equation}
Y({\bf q},a)= \int_{a_s}^{a}{\frac{a^{\frac{3}{2}}da}
     {q_{x}^{2} +[q_{y}-q'(a)]^{2}}} \quad.  \label {eq:14}
\end{equation}

We see that $Y$ scales like $a^{5/2}$ when $a \rightarrow\infty$.
In this case it is sufficient to keep only the leading term in $a$
for the  decaying mode:
\begin{equation}
a^{-3/2}J_{x} \simeq\frac{18t_{0}^2}{25}\lambda G  a \frac{q_x}
{q_{x}^{2} +(q_{y}-q_{i})^{2}} \quad, \label{eq:15}
\end{equation}
It is worth mentioning that the above equation follows from (\ref{eq:13}) and
(\ref{eq:14}) by neglecting terms of order of  $a^{1/2}$ (and lower). Within
such an approximation we may neglect the $a$ dependence of $q'$ in the  $ln$
and $tan^{-1}$ terms in equation (\ref{eq:12}), because its
contribution is of the same order as
terms neglected in (\ref{eq:15}).

Using Equations (\ref{eq:10})--(\ref{eq:15}) it is easy to shown that the
turn-around
condition (which can be evaluated at any redshift $z$) is given by
\begin{eqnarray}
q_{x}^2&+&(q_{y}-q_{i})^{2}=b(t)\left\lbrace ln\left(\frac{a}{a_s}\right)
+ ln\left(\frac{q_{x}^{2}+ (q_{y}-q_{i})^{2}}
{q_{x}^{2}+(q_{y}-q_s)^{2}}\right)+\frac{1}{10}+ \right.\nonumber \\ & &\left.
+2\frac{q_{y}-q_i}{q_x}\left\lbrack tan^{-1}\left(\frac{q_{y}-q_i}{q_x}\right)
 -tan^{-1}\left(\frac{q_y-q_s}{q_x}\right)\right]\right\rbrace \,
,\label{eq:16}
\end{eqnarray}
with
\begin{equation}
b(t) \equiv\frac{18t_{0}^2}{5} \lambda Ga \, . \label{eq:17}
\end{equation}
Note that the turn-around surface has the form of a tube whose perimeter is
defined by
(\ref{eq:16}) and whose length is proportional to the comoving distance
corresponding to the Hubble radius at time $t_i$. Thus, the comoving length
$l_c(t_i)$ of the string at $t_i$ is given by
\begin{equation}
l_c(t_i) = \xi (1 + z(t_i)) t_i\, ,
\label{eq:length}
\end{equation}
where $\xi$ is a constant which determines the curvature radius of the long
string network relative to $t$ and whose value will be discussed in Section 6.
For CDM the initial conditions are $q_s=0$, $a_s=a_i$ with the right hand side
of (\ref{eq:16}) reducing asymptotically to
\begin{eqnarray}
Q_{CDM}^{2}&=&b(t)\left\lbrace ln\left(\frac{a}{a_i}\right)
+
ln\left(\frac{q_{x}^{2}+(q_{y}-q_{i})^{2}}{q_{x}^{2}+q_{y}^{2}}\right)+\frac{1}{10}+\right. \nonumber\\ & & \left.
+2\frac{q_{y}-q_i}{q_x}\left\lbrack tan^{-1}\left(\frac{q_{y}-q_i}{q_x}\right)
 -tan^{-1}\left(\frac{q_y}{q_x}\right)\right]\right\rbrace \, ,\label{eq:18}
\end{eqnarray}
where we defined $Q^2_{CDM}:=q_{x}^2+(q_{y}-q_{i})^{2}$.

The perimeters of the turn-around surfaces at various redshifts are depicted in
Figure 1.
Unlike in the static case, the string motion along the $y$ axis gives rise to a
suppression of the turn-around distance along the $x$ axis. For moving loops,
the
same effect has been  established
by Bertschinger\cite{EB87}.
It is easy to understand the shape of these curves. The
filament starts its motion at $|{\bf q}'|=0$ with a high velocity $u_i$. Due
to the expansion of the Universe, the peculiar velocity decreases at later
times
as $a(t)^{-1}$. As a consequence, the seed spends much less time in the
region, say, $q_y < 0.9q_i$ than in the region $q_y \sim q_i$. The integrated
action of gravity is hence larger in the second region. This explains why the
turn-around distance is greater near the final position of the filament (see
also Figure 2).
It should be noted, however, that if the filament
is replaced by
a string loop, the turn-around surface will be
thinner nearby the end of the motion. This can be
understood by the fact that the gravitational pull for loops ($F\sim 1/r^2$) is
less efficient at  large distances than for filaments ($F\sim 1/r$).

\begin{center}
\bf Figure 1
\end{center}
\begin{figure}
\centerline{
\psfig{figure=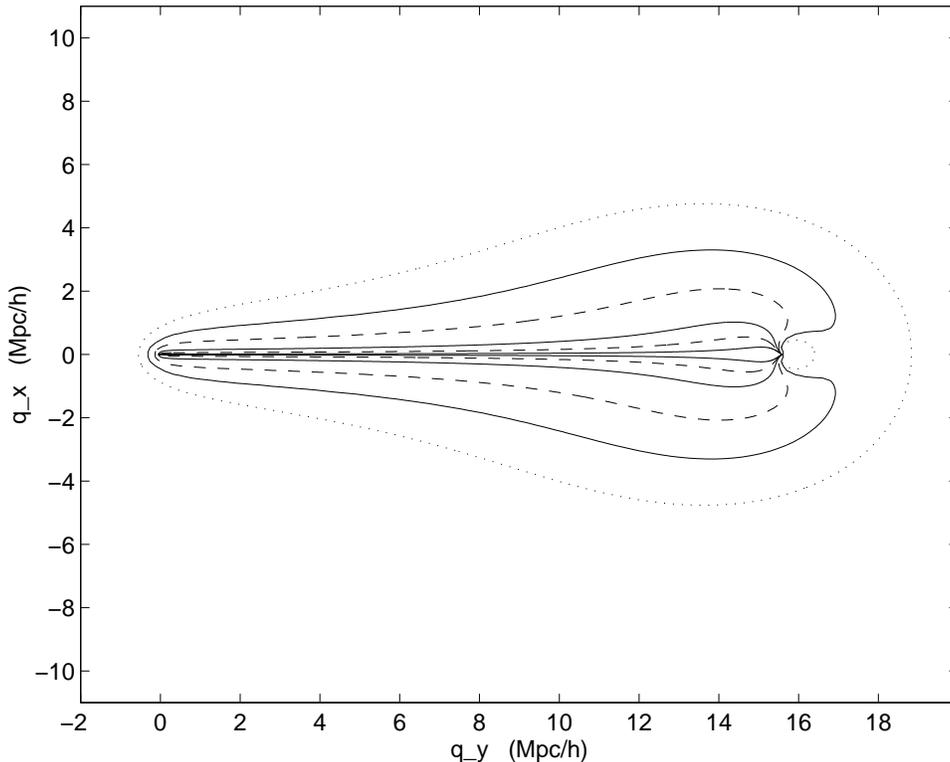,height=4in}}
\caption{The comoving perimeter of the turn-around surface for the moving
filament, with initial velocity $u_i=.2$ and CDM. The curves are for
the redshifts $z = 200$ (innermost contour), $50$, $20$, $5$, $3$, and $1$
(outermost contour). For $z < 2.1$ the turn-around consists of two contours.}
\end{figure}

It is  worth mentioning that the accretion is also suppressed (on small
scales, $q_x \approx 0$)
near ${\bf q} =(0,q_i)$, as happens
with the small scale suppression due to free streaming. This is shown in
Figure 2 where the turn-around redshift is plotted as a function of $q_{x}$
for $q_{y}= q_i/2$, and for $q_{y}=q_i$ where such a suppression is observed
(dotted line and
continuous line). These curves look like the case of a static filament in HDM
models\cite{Aguirre},
 however, here the suppression at a fixed time $t\gg t_i$ occurs only on scales
much smaller than
 the suppression caused by the free streaming. This is readily seen by
rewriting equation (\ref{eq:18}) for $q_y=q_i$
\begin{equation}
q_x^2= q_{x,st}^2 + b(t)\, ln\left(\frac{q_x^2}{q_x^2+q_i^2}\right)\, ,
\label{eq:19}
\end{equation}
where $q_{x,st}$ is the turn-around comoving position (the $x$ component) in
the case of a stationary filament. For small values of $|q_x|$,
the $ln$ term on the right hand side of
(\ref{eq:19}) contributes with a very large negative
value (the $tan^{-1}$ terms are finite even for $q_y\neq q_i$), thereby
implying that turn-around is
possible in that region just for very large times (compared to $t_i$).
This suppression is also a consequence of the peculiar motion of the string
whose
gravitational force on a cold dark particle acts in different (though
correlated) directions at different times, which is not the case on large
scales. Such a  small scale
suppression does not occur in the case of loops\cite{EB87}. This can also be
explained by taking into account the gravitational action of each defect.

\begin{center}
\bf Figure 2
\end{center}
\begin{figure}
\centerline{
\psfig{figure=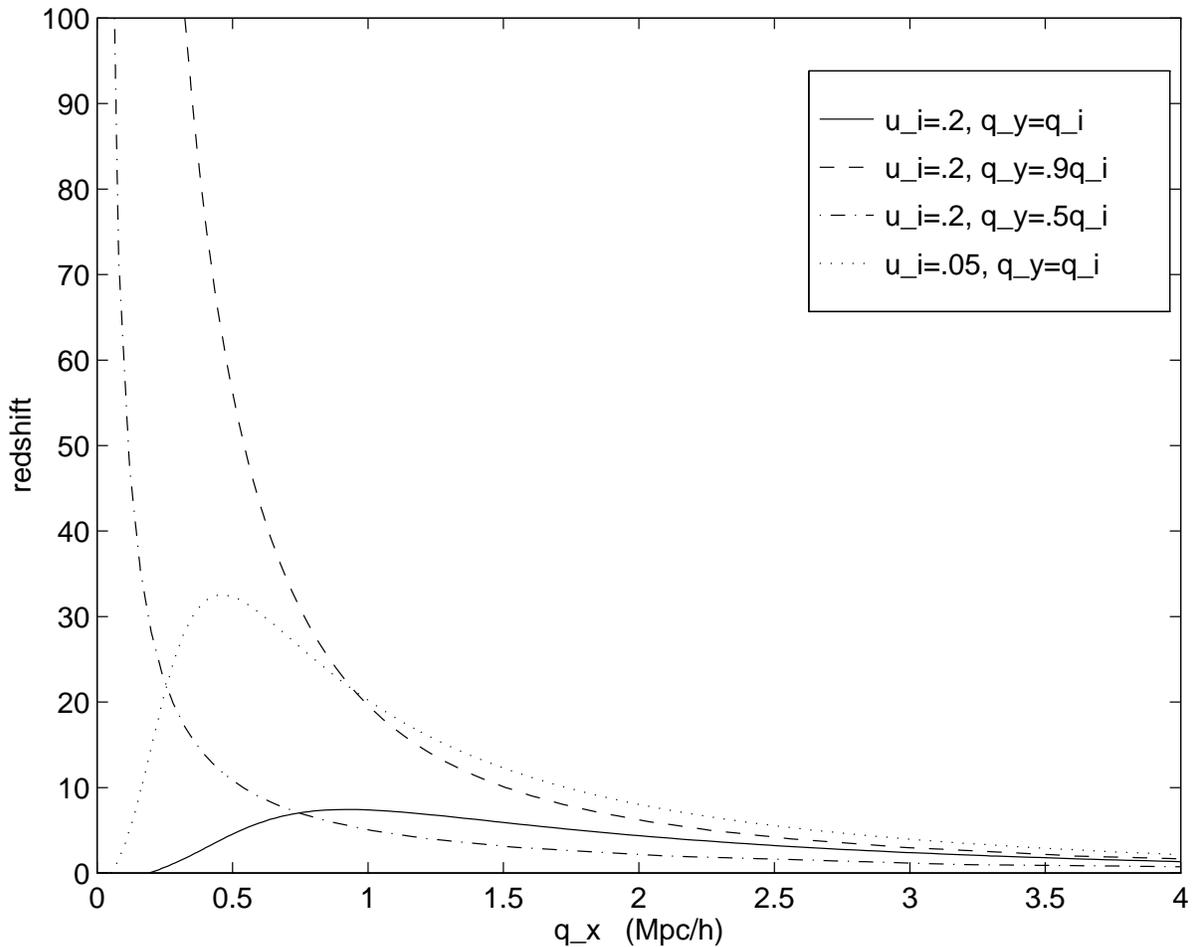,height=5in}}
\caption{Redshift versus $q_x$ for different values of $q_y$ and $u_i$.}
\end{figure}

The overall effect of the string motion is to
distort the cross-section of the turn-around
surface (a circle in the
stationary case). However, the suppression along the $x$ axis is more than
compensated by the growth along the $y$ axis (see Figure 1). In fact, the
suppression along $x$                                        is only
logarithmic in the string velocity (it goes as $ln(1/q_i)$) while the length of
the wake along the $y$ axis increases linearly (proportional to $q_i$).
In contrast, the total amount of CDM accreted by a string loop is independent
of the velocity (see Refs. \cite{Rees} and \cite{EB87}). Figure 3 shows the
dependence of the maximum value $q_{nl,x}(q_y)$ as a function of the velocity
of the seed.

\begin{center}
\bf Figure 3
\end{center}
\begin{figure}
\centerline{
\psfig{figure=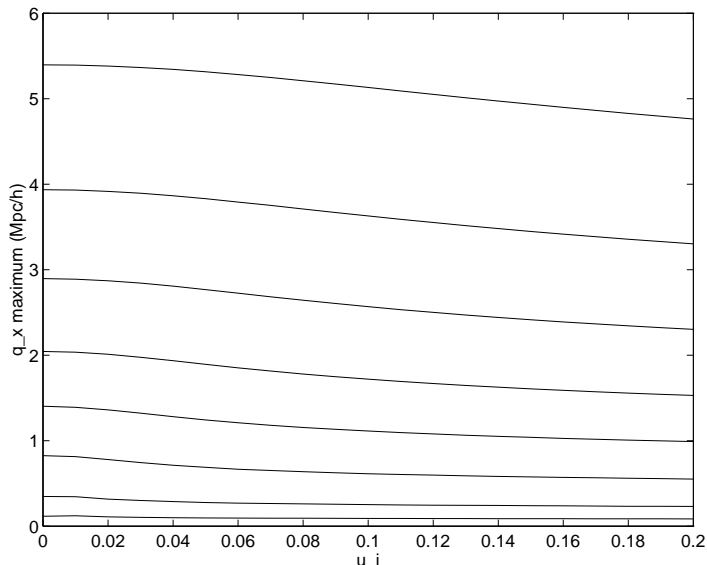,height=3in}}
\caption{Variation of the maximum of $q_{nl,x}(q_y)$ as a function
of the velocity
of the seed. The plots are for $ z=1000$ (lowermost curve), $200$, $50$, $20$,
$10$, $5$,
$2.5$, and $1$ (uppermost curve). The difference for $q_{nl,x}(q_y,u_i)$
between
$u_i=0$ and $u_i= .2$  falls from $25\%$  for $z=1000$ to $11\%$ for $z=1$.}
\end{figure}

We have calculated numerically the total nonlinear mass (i. e. the mass inside
the turn-around surface) as a function of the redshift. A reasonable
approximation can be obtained by assuming that the area inside the
turn-around curve defined by (\ref{eq:18}) is given by
\begin{equation}
A = \pi Q_{CDM}^2 + q_i \frac{l_x}{2}\, ,
\label{eq:area}
\end{equation}
where $l_x$ is the the comoving width of
the turn-around volume at its thickest point (near $q_y=q_i$).  The second term
in (\ref{eq:area}) corresponds to the area of an isosceles triangle whose base
is $l_x$ and height is $q_i$.
This equation in fact yields a lower limit for the total
mass that has gone nonlinear. Since a lower bound to $\Omega_{nl}$
will be derived in Section 5, it is appropriate at this point to obtain a lower
limit to the accreted mass.

The second term in (\ref{eq:area}) is proportional to $u_i$, the initial
velocity of the string, whereas the first term in the limit $u_i = 0$ reduces
to the results of Ref. \cite{Aguirre} for the static case. Comparing the two
terms, we see that the second term in (\ref{eq:area}), stemming from the motion
of the string, gives the dominant contribution if
\begin{equation}
{{2 \pi} \over {15 u_i}} (\lambda G)_6^{1/2} (z(t) + 1)^{-1/2} h \Omega^{1/2} <
1,
\label{eq:comparison}
\end{equation}
where $(\lambda G)_6$ is the value of $\lambda G$ in units of $10^{-6}$. For
example, we see that if $t_i = t_{eq}$, $(\lambda G)_6 = 0.5$, $h = 1/2$,
$\Omega = 1$ and
$u_i = 0.2$, then the two terms are comparable at redshift 0, but the
contribution originating from the second term in (\ref{eq:area}), the
contribution specifically due to the motion of the string, dominates for higher
redshifts.

The total mass accreted can be written as (see Equation (\ref{eq:length}))
\begin{equation}
M(z,u_i) = \xi \rho_0t_i(z_i + 1) A(z,u_i) \, , \label{eq:20}
\end{equation}
where $(z_i + 1)=(t_0/t_i)^{2/3}$ and $A(z,u_i)$ is the area given by
(\ref{eq:area}). As a consistency check, we note that
taking the static limit ($q_i=0$) in the
above equations, the results
of Ref. \cite{Aguirre} are readily recovered. For the values of the parameters
used above,
Equation (\ref{eq:20}) yields:
$M \simeq 5 \xi \times 10^{12}M_\odot A(z,u_i)/Mpc^2$.

To proceed further, it proves convenient
to summarize the main properties of the
turn-around curve described by (\ref{eq:18}). For CDM we have checked
numerically:

 (i) The comoving area grows linearly with the scale factor (as a
consequence of the fact that $ln(a/a_i)$
is a slowly varying function of $a$) for $a\gg a_i$.

 (ii) The area inside the turn-around curve grows almost linearly
with the velocity of the filament $u_i$ for $u_i > .05$ and is practically
constant for $u_i < .05$ at redshifts under consideration (because in the
latter case the term in (\ref{eq:area}) independent of $u_i$ dominates).

 (iii) Due to the motion of the filament, the accretion on
large scales is strongly suppressed near the comoving initial position of the
string (because the gravitational line source is far away for most of the
time), whereas growth on small scales is suppressed
at the end, when $q'\sim q_i$.

 (iv) Using $u_i=0.2$ and $t_i = t_{eq}$ we obtain
at the redshift
$z=4$ a turn-around curve whose length is $l\approx q_i
\simeq 8h^{-2} Mpc$ and whose width in the thickest part $l_x$ is about
$2 h^{-1} (\lambda G)_6^{1/2} Mpc$. The area $A$ inside
the curve is approximately $16 h^{-3} (\lambda G)_6^{1/2} Mpc^2$ which
corresponds to a mass per unit length of
$M= \xi A\rho_0 t_{eq}z_{eq} \simeq 50\xi\times 10^{12} h^{-1} (\lambda
G)_6^{1/2} M_\odot$.

\hspace{.3in}

\section{Accretion of Hot Dark Matter}

 Now we turn to the accretion of hot dark matter. The basic equations
(\ref{eq:12} - \ref{eq:length}) which describe accretion onto a moving
Newtonian line source are applicable.

We will deal with the ambiguity in the definition of $t_s(q)$ mentioned in
Section 2 in the following way: Since we are only interested
in the motion of dark matter in the direction perpendicular to the seed
velocity, we may replace the initial condition (\ref{eq:07}) by
\begin{equation}
t_s =\frac{t_0\lambda_{0}^3}{|q_x|^{3}}\,. \label{eq:08}
\end{equation}
This, however, seems to
overestimate the free streaming effect, since it completely erases the
accretion in the region near the initial location of the string (an improved
analysis using the Boltzmann equation is in progress and will be reported
elsewhere). In the following we will adopt
(\ref{eq:08}) as a generalization of the  condition used for static seeds which
is a good approximation for the real free streaming effect
\cite{PBS90}. The
quantities we are going to obtain in this case, such as the total mass
accreted, are lower bounds to the exact ones.

Since we must have $t_s \geq t_i$, condition (\ref{eq:08}) can be
applied only if $|q_x| \leq v_i t_i (z_i + 1)$, otherwise the
condition $t_s=t_i$ is to be used even for HDM.

Thus, only a few small changes must be made: In the turn-around conditions we
substitute $a_{s} = \frac{\lambda_{0}^2}{|q_x|^2}$. Then for HDM, equation
(\ref{eq:16}) may be cast as
\begin{eqnarray}
 q_{x}^2&+&(q_{y}-q_{i})^{2}= Q_{CDM}^2  + b(t)\left\lbrace
 ln\left(\frac{q_x^2}{\lambda_i^2}\right)+ln\left(\frac{q_{x}^{2}+q_{y}^{2}}
 {q_{x}^{2}+(q_{y}-q_s)^{2}}\right) \right.\nonumber \\ & &\left.
-2\frac{q_{y}-q_i}{q_x}\left\lbrack tan^{-1}\left(\frac{q_{y}-q_s}{q_x}\right)
 -tan^{-1}\left(\frac{q_y}{q_x}\right)\right]\right\rbrace \, . \label{eq:21}
\end{eqnarray}
where $q_s= q_i(1-|q_x|/\lambda_i)$, and $\lambda_i=\lambda_0/\sqrt{a_i} =
 \lambda_J(t_i)$.

As shown in Figure 4, the turn-around curve (\ref{eq:21}) is composed of two
identical disconnected closed curves,
one for negative values of $q_x$ and the other for positive values of $q_x$.
It is apparent from the figure that the nonlinearities on small scales are
highly suppressed by the free streaming condition - see (\ref{eq:07}).

\begin{center}
\bf Figure 4
\end{center}
\begin{figure}
\centerline{
\psfig{figure=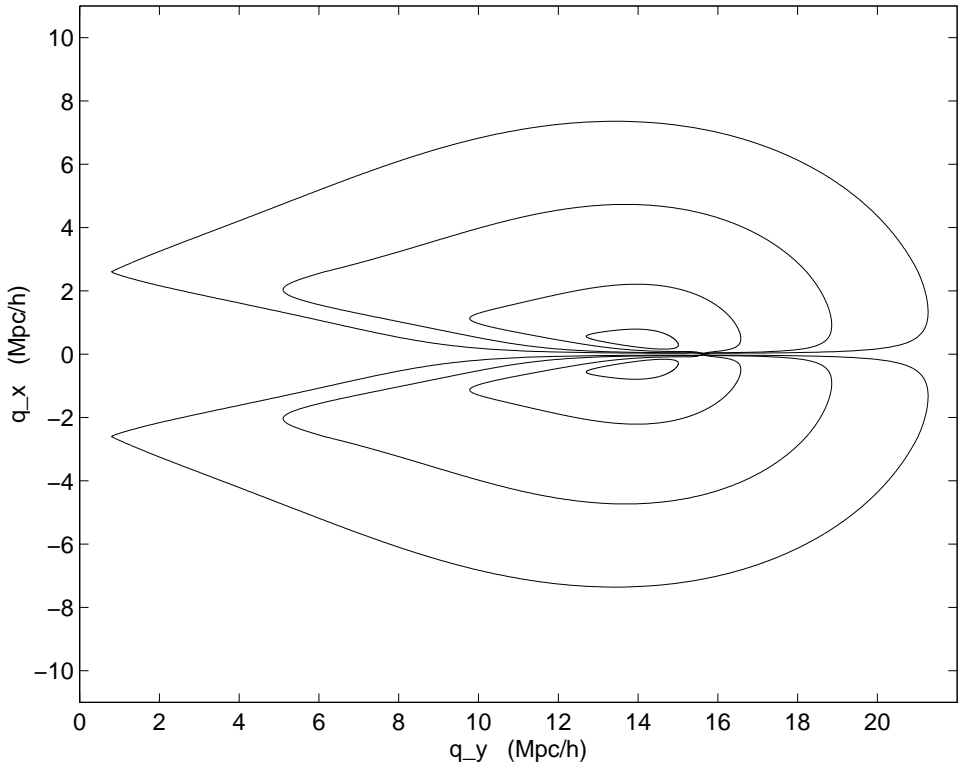,height=4in}}
\caption{The same as Figure 1, but now for HDM with velocity $v_{eq} =.1$.
Contours (two sheets each one) are drawn for $z = 20$ (inner contour), $5$,
$1$, and $0$ (outer contour).}
\end{figure}

In order to compare to other cases let us rewrite (\ref{eq:21}) for $q_y=q_i$
\begin{eqnarray}
 q_{x}^2 = q_{x,CDM}^2+ b(t)\left\lbrace
ln\left(\frac{q_x^2}{\lambda_i^2}\right)+ln\left(\frac{q_{x}^{2}+q_{i}^{2}}
{q_{x}^{2}+q_i^2{|q_x|}{\lambda_i}^{-1}}\right)\right\rbrace \, . \label{eq:22}
\end{eqnarray}

Recalling that equation (\ref{eq:22}) applies for $|q_x|\leq \lambda_i$
(for larger values of $|q_x|$ the CDM result is valid) we see that the first
$ln$
term on the right hand side of (\ref{eq:22}) is negative while the
second one is positive. At early times, when $q_x^2 << q_i^2$ those two
terms reduce to $ln(|q_x|/\lambda_i)$ and a numerical analysis shows that
for $v_i \leq u_i$ the ratio of this term to $q_{x,CDM}^2$ is very small. This
means that for a rapidly
moving filament the suppression of the turn-around near $q_y=q_i$ is governed
by the seed motion and is the same for both CDM and HDM. For $v_i > u_i$ the
suppression due to the free streaming is important also for $q_y=q_i$. These
features can be seen from Figures 2 and 5, graphs which plot the redshift
corresponding to the onset of nonlinearity as a function of
$q_x$ for different velocities of the string. For instance, for the value $u_i
= 0.2$, the curves
for $q_y=q_i$ in the cases of CDM (Figure 2) and HDM (Figure 5) are essentially
the same, whereas for the value $u_i = 0.05$, the HDM curve is substantially
lower at small values of $q_x$.
Figure 5 also shows that there are no nonlinear structures for
redshifts greater than about 90.

\begin{center}
\bf Figure 5
\end{center}
\begin{figure}
\centerline{
\psfig{figure=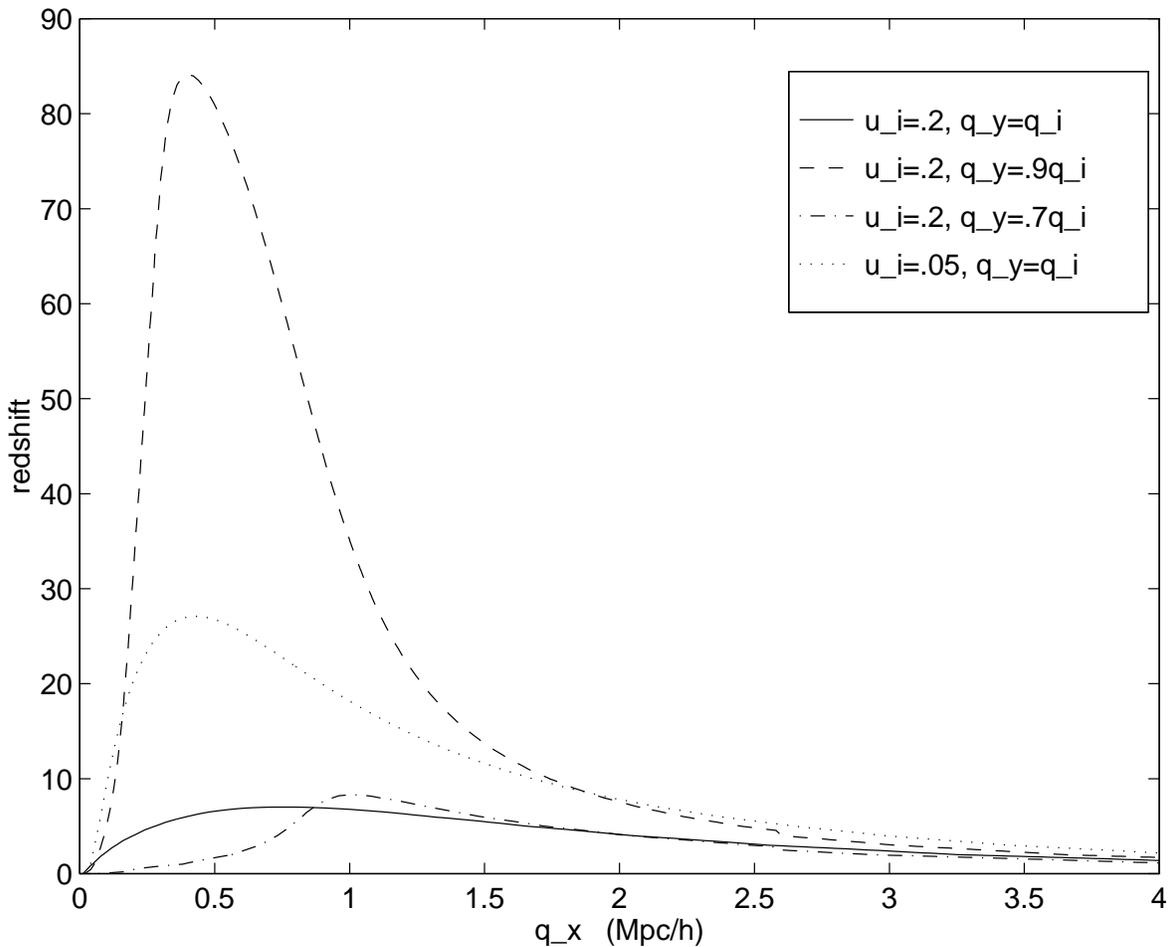,height=5in}}
\caption{Same as figure 2, now for HDM with velocity $v_{eq}=.1$.}
\end{figure}

We know that free streaming completely erases the nonlinearities
on scales smaller than the free streaming length. Equation (\ref{eq:21})
shows this behavior as a consequence of
the term $ln(q_x^2/\lambda_i^2)$, which is independent of $q_y$ and for small
$|q_x|$ is highly negative implying that turn-around can only occur
at late times.
As we have seen in the previous section, for $q_y < q_i$ the $tan^{-1}$ terms
are also important in determining the turn-around on
large scales (they are finite even for $q_x \longrightarrow 0$). Then, the
presence of the new $tan^{-1}$ term in (\ref{eq:22}) will not significantly
change the results, and it follows that for a moving filament
with HDM, the wake is strongly erased in the thinnest region near the
starting position of the seed (see Figure 4).

The total nonlinear mass accreted in the HDM
model can be estimated by considering
all the mass inside and between the two disjoint curves (see Figure 4). We then
approximate the turn-around curve by a trapezoid. Such a figure figure is
obtained
when the height $q_i$ of the isosceles triangle considered in the CDM
case is reduced to $l_y$, thereby accounting for the free streaming
suppression. By comparing Figures 1 and 4 it can be seen that the difference in
the nonlinear mass between HDM and CDM is large at high redshift and decreases
at later times.
The length $l_y$
of the turn-around curve along the $y$ axis depends on the redshift and tends
to $q_i$ for very late times (e.g. for $z<2$ if $t_i=t_{eq}$). According to
Figure 4, a simple (underestimating) approximation is to take
$l_y=q_i-q_y(\lambda_i)$, where $q_y(\lambda_i)$ is given by the intersection
between the line $q_x=-\lambda_i q_y/q_i +\lambda_i$ and the approximated
turn-around curve
 for CDM (see the comments below (\ref{eq:area})). The result is
$l_y=q_i-\lambda_i q_i/(\lambda_i+l_x/2)$.
  Hence, the area of the nonlinear region can be estimated as
\begin{equation}
A_{HDM}=\pi Q_{HDM}^2 + q_i \frac{l_x}{2}\left(1-
\frac{\lambda_i^2}{(\lambda_i+l_x/2)^2}\right) \, ,
\label{eq:areahdm} \end{equation}
where $Q_{HDM}^2$ stands for the right hand side of (\ref{eq:22}).
The first term is analogous to the one introduced in the CDM case, the first
term in
equation (\ref{eq:area}), which represents the accretion onto a static filament
(at the comoving position $q_i$) with HDM. For $t$ sufficiently greater than
$t_i$ the second term
in (\ref{eq:areahdm}) dominates if the same condition for CDM, Equation
(\ref{eq:comparison}), is
satisfied. In such a case we can parameterize the result as follows:

\begin{equation}
M_{HDM}(t, t_i, u_i) = f(t, t_i, u_i) M_{CDM}(t, t_i, u_i),
\label{eq:ratio}
\end{equation}
where $f(t, t_i, u_i)$ is the function which parameterizes the mass suppression
in the case of HDM. For large $t$, the result with HDM asymptotically
approaches the CDM result. Hence
\begin{equation}
lim_{t \rightarrow \infty} f(t, t_i, u_i) = 1.
\label{eq:limit}
\end{equation}
The numerical analysis shows that for the cosmological and cosmic string
parameters used (see Section 3),
at a redshift of
$z=10$, $f(t, t_{eq}, u_i) \approx 0.5$, and for $z=4$, $f(t, t_{eq}, u_i)
\approx 0.8$.  The function
$f(t, t_i, u_i)= 1- \frac{\lambda_i^2}{(\lambda_i+l_x/2)^2}$
used in (\ref{eq:areahdm}) furnishes the following approximate values: $f=0.5$,
$f= 0.85$, and $f=0.96$ respectively for
$z=10$, $z=4$ and $z=0$. We have taken $l_x=2|q_x|$ with $q_x$ given by
(\ref{eq:area}) (see following discussion).

\section{The Nonlinear Density Parameter  $\Omega_{{n}{l}}$}

To illustrate the results of the previous sections, let us now compute the
fraction $\Omega_{nl}$ of the critical density accreted onto filaments
both for CDM and HDM models. Naturally, by
considering only the accretion onto filamentary
structures we will be underestimating the total contribution to $\Omega_{nl}$
due to the cosmic string network. Let us first
consider the case where the initial string velocities $u_i$ and times $t$ are
such that the second term in (\ref{eq:area}), the term specific to the motion
of the string,
dominates. This is the case of primary interest to us, since in the other
limit($q_{i}=0$) the  results for static filaments are recovered (see
eqs.(20) and (26)).

As argued earlier (see Section 3), to estimate $\Omega_{nl}$ in the case of
moving filaments with CDM, one may approximate the turn-around curve by an
isosceles triangle whose height and base are respectively $l_y=q_i$ and
$l_x$, where $l_x$ is the width of the thickest region
of the wake. A good approximation for the width is $2|q_x|$, where
$q_x$ is given by (\ref{eq:19}) without the term $ln[q_x^2/(q_x^2+q_i^2)]$.
This term leads to a suppression of small scales, whereas for large scales it
is small in
comparison to $ln(a/a_i)$. In this way, by considering
the mass of the wake for $0\leq q_y\leq q_i$,
such a term may be safely neglected. The resulting expression for $q_x$ is then
the same as for a static filament located at $q_i$ since the initial time
$t_i$. Also, the condition
that the string is not near $q_i$ for all times after $t_i$, can be effectively
described
replacing  $t_i$ by the rescaled time $t_{eff}= \alpha t_i$, where $0 <
\alpha\leq 1$.
The time $t_{eff}$ can be thought as a sort
of initial time for which the Newtonian
line starts to act effectively as a static
source for accretion
near $q_i$.  The value of
$\alpha$ may be fixed by comparing the exact value of $q_x$ obtained
numerically with the approximation given by $q_{x,static}(t,\alpha t_i)$.
As can be easily seen, for the values of
the parameters considered here one has $\alpha \approx 10^{-1}$. In the limit
$\alpha \rightarrow 1$, the relative error for sufficiently large times may be
written as
$ln(\alpha)/ln(t/t_i)$, and for $t_i=t_{eq}$, it
is about $15\%$ at the present time (see Figure 3 for comparison).

The total mass accreted onto the filament during an arbitrary time interval
$\Delta t= t-t_i$ ($t_i\leq t\leq t_o$), is the mass
contained in the Lagrangian volume inside the turn-around surface. Since the
length of the turn-around tube is proportional to the
Hubble radius at time $t_i$, the accreted mass is given by (see Section 3)
\begin{equation}
M(t,t_i) =
\xi\frac{u_i}{2\pi}\sqrt{\frac{18\lambda}{5G}}t_i\left(\frac{t}{t_i}\right)^{1/3}
\sqrt{ln\left(\frac{t}{t_i}\right) +\frac{3}{20}}  \, . \label{eq:23}
\end{equation}
This is the total mass that has gone nonlinear at time $t>t_i$ onto a Newtonian
line formed at time $t_i$.
Note that the nonlinear mass grows as $t^{1/3}$ and not as $t^{2/3}$ as in the
linear theory for static seeds. The difference comes from the fact that the
length along the direction of string
motion depends only on the velocity of the seed,
and is, to the first order of approximation,  time independent. As expected,
for slowly moving strings, the first term in (\ref{eq:area}) dominates, and the
usual linear perturbation theory growth of the total nonlinear mass is
recovered. This can also
be seen by using (\ref{eq:comparison}) to determine the lower cutoff velocity
when (\ref{eq:23}) is applicable, and substituting in (\ref{eq:23}) for
the velocity $u_i$. This yields the extra factor of $(z(t) + 1)^{-1/2}$.

{F}rom the scaling solution of the string network we know that the number of
long strings per physical volume at time $t$ created
in the interval between $t_i$ and $t_i+dt_i$
assume the form
\begin{equation}
n(t,t_i)dt_i =\frac{\tilde{\nu}}{t_i^{2}t^{2}}dt_i \, ,
\label{eq:distr}
\end{equation}
where $\tilde{\nu}$ is a constant which gives the number of long string
segments passing through a volume $t^3$ at any given time (this number is
independent of time as a consequence of string scaling). Using this number
density  we  obtain
the fraction of the critical density of objects formed by accretion of CDM onto
moving filaments in the interval between $t_{eq}$ and $t$  as
\begin{equation}
\Omega_{nl}(t)=6\pi G t^2 \int_{t_{eq}}^{t}{n(t,t_i)M(t,t_i) dt_i} \, ,
\label{eq:omega}
\end{equation}
where $M(t,t_i)$ is given by (\ref{eq:23}).

Then by direct integration of (\ref{eq:omega}) we obtain to a first
approximation
\begin{eqnarray}
&\Omega_{nl}(t) &\approx 9\xi\tilde{\nu}u_i\sqrt{\frac{18\lambda G}{5}}
\sqrt{\ln\left(\frac{t}{t_{eq}}\right)}\left(\frac{t}{t_{eq}}\right)^{1/3}
\nonumber \\& &\approx 2\xi\tilde{\nu} (u_i)_{0.2} (\lambda G)_6^{1/2} h(z(t) +
1)^{-1/2} \, ,
\label{eq:24}
\end{eqnarray}
where $(u_i)_{0.2}$ is the value of $u_i$ in units of $0.2$. It follows
that for $u_i = 0.2$ and $(\lambda G)_6 = 0.5$ the value of $\Omega_{nl}(t_0)$
at the present time $t_0$ is about $2\xi\tilde{\nu}$. Note that the above
result is valid for values of $u_i$ for which (\ref{eq:comparison}) is
satisfied. For
smaller initial string velocities,
by solving (\ref{eq:comparison}) for $u_i$ and inserting in   (\ref{eq:24})),
we obtain in the static limit
\begin{equation}
\Omega_{nl, st}(t) =\frac{18}
{5}\pi\lambda G \tilde{\nu}\left(\frac{t}{t_{eq}}\right)^{2/3}
 ln\left(\frac{t}{t_{eq}}\right) \approx \xi \tilde{\nu} {{2 \pi} \over 3}
(\lambda G)_6 (z(t) + 1)^{-1} h^2\, . \label{eq:24b}
\end{equation}

{F}or HDM we have approximated the turn-around curve by a trapezoid (see the
discussion below
equation (\ref{eq:22})). The velocity depending area $A_{HDM}$ inside such a
curve is given by
\begin{equation}
A_{HDM}= f(t,t_i,u_i)A_{CDM} =  q_i \frac{l_x}{2}
\left(1- \frac{\lambda_i^2}{(\lambda_i+l_x/2)^2}\right) \, ,
\label{eq:areahdm1}
\end{equation}

As mentioned before, this area includes
the holes inside and between
the curves of Figure 4. This means that
all $ln$ terms in (\ref{eq:22}) are neglected. As a matter of fact, the
suppression terms
due to free streaming are important only for small scales which have
been by the above approximation included in $A_{HDM}$.
Besides, the term in (\ref{eq:22}) arising from the motion
of the string may also be neglected by the same reason as discussed earlier.
As a consequence, the  expression of $q_x$ for
HDM reduces to the same one derived in the CDM model.

The value of $\Omega_{nl}$ in the case of HDM can be estimated by following
 (\ref{eq:omega}):
\begin{equation}
\Omega_{nl}^{HDM} = 6\pi G t^2\int_{t_{eq}}^t n(t,t_i) f(t,t_i,u_i)
M_{CDM}(t,t_i,u_i) dt_i\, .
\label{eq:26}
\end{equation}
Since the integral is dominated at the lower end, we can approximate the result
as
\begin{eqnarray}
&\Omega_{nl}^{HDM} (t) &\simeq f(t, t_{eq}, u_i) \Omega_{nl}^{CDM} (t)\,
\nonumber\\
& &\simeq\left(1-
\frac{\lambda_{eq}^2}{[\lambda_{eq}+l_x(t,t_{eq})/2]^2}\right)
\Omega_{nl}^{CDM}\, ,
\label{eq:27} \end{eqnarray}
where, as we have seen above, $f(t,t_{eq},u_i)$ grows from $f\simeq 0$ at
$t=t_{eq}$
 to $f\simeq 0.96$ at $t=t_0$.

    {F}or the sake of completeness, let us now write  the contributions from
the first terms in equations
(\ref{eq:area}) and (\ref{eq:areahdm}) to the fraction of nonlinear mass. As
discussed earlier,
these terms correspond to the
fraction of the accreted mass onto static filaments. Following the previous
procedure we get
\begin{equation}
\Omega_{nl,st}^{CDM} \simeq
\frac{18}{5}\pi\xi\tilde{\nu}\lambda G\left(\frac{t}{\alpha
t_{eq}}\right)^{2/3}
 ln\left(\frac{t}{\alpha t_{eq}}\right) \, , \label{omegast1}
\end{equation}
(compare with (\ref{eq:24b})) and for HDM
\begin{equation}
\Omega_{nl,st}^{HDM} \simeq (1- \chi )\Omega_{nl,st}^
{CDM} \, ,\label{omegast2}
\end{equation}
where $1-\chi$ is the suppression factor due to free streaming. $\chi =\chi
(t,t_{eq})$ falls from
$\chi =1$ for $t=t_{eq}$ to $\chi \approx 0$ at $t=t_0$.  Notice that
$\Omega_{nl,st}$ varies
with $(z+1)^{-1}$ while the contribution due to the string motion, equations
(\ref{eq:24}) and
(\ref{eq:27}), goes as $(z+1)^{-1/2}$. This fact explains why $\Omega_{nl,st}$
is important
just for small $z$, when compared to the terms depending on the seed velocity.

These results show  that even with hot dark matter (given the string parameters
used above), accretion onto filaments can easily lead to $\Omega_{nl} = 1$ at
the present time.

\section{Conclusions}

We have studied the accretion of hot and cold dark matter onto moving
gravitational line sources and applied the results to the cosmic string
filament model of structure formation. Our main result is that even for initial
relativistic motion of the strings, nonlinear structures can develop at
redshifts close to 100 if the dark matter is hot (a point not realized in
earlier work \cite{VV91}). Thus, string filaments are much more efficient at
clustering hot dark matter at early times than string wakes.

The total fraction $\Omega_{nl}$ of nonlinear mass depends quite sensitively on
the string parameters used, i.e. on the initial string velocity $u_i$, on the
strength $\lambda G$ of the Newtonian potential, and on the values of the
constants $\xi$ and $\tilde{\nu}$ which determine the curvature radius and
number of strings in the scaling solution. However, the uncertainties partially
cancel. For example, the product $\tilde{\nu} \xi$ is a measure of the total
mass density in strings and can in principle be determined from numerical
simulations of cosmic string evolution. Also, as $\lambda G$ increases, the
value of $u_i$ will decrease, but in this case the cancelation is only partial
since $u_i$ cannot exceed 1 and hence in the limit $\lambda G \rightarrow 0$
the filamentary accretion will disappear (and the only effect of the strings
will be the wakes caused by the velocity perturbations).

According to the numerical simulations of Allen and Shellard \cite{CSsimuls}
(which, however, are based on strings obeying the Nambu-Goto action and hence
not directly applicable to strings with a large amount of small-scale
structure), the value of $\xi \tilde{\nu}$ is about 15 in the radiation
dominated epoch and slightly less during matter domination. Taking the value
$(\lambda G)_6 = 0.5$ which corresponds to the largest possible effect of
small-scale structure for GUT scale strings, we find that even for hot dark
matter, all of the mass in the Universe has gone nonlinear by a redshift of
about 5. This implies that the cosmic string model - provided the small-scale
structure dominates - is easily compatible with the cosmological constraints
from both high redshift quasar \cite{QSO} and damped Lyman alpha system
\cite{DLAS} abundances.

It is interesting to compare the results derived here with those from the
cosmic string wake \cite{PBS90} and loop \cite{RM96} scenarios. If the dark
matter is hot, wakes do not produce any nonlinearities until a redshift of
close to 1. Loops do produce nonlinear objects at high redshifts, but objects
of mass
$10^{12} M_\odot$ only form at a redshift of about 4. Hence, we conclude that
string filaments are the most efficient way of forming high redshift massive
nonlinear structures in the cosmic string model. Unfortunately, it is not known
how much small-scale structure cosmic strings are endowed with. The cosmic
string evolution simulations at present do not have the required range to
resolve this issue, and analytical analyses are needed to address this
question.
In light of the sensitive dependence of high redshift structure formation to
the amount of small-scale structure, it is crucial to improve our understanding
of the small-scale structure on strings.

\centerline{\bf Acknowledgments}

The authors are grateful to R. Moessner
and M. Parry for useful discussions. This work is supported in part by the
Conselho Nacional de
Desenvolvimento Cient\'{i}fico e
Tecnol\'{o}gico-
CNPQ(Brazilian Research Agency), and by the US Department of Energy under
contract DE-FG0291ER40688, Task A .

\end{document}